\documentclass[english,amsmath,amssymb,superscriptaddress,nofootinbib,twocolumn]{revtex4-1}
\usepackage[LGR,T1]{fontenc}
\usepackage[latin9]{inputenc}
\usepackage{amsmath}
\usepackage{amssymb}
\usepackage{graphicx}

\makeatletter

\ProvideTextCommand{\~}{LGR}[1]{\char126#1}

\usepackage[normalem]{ulem}

\usepackage{color}

\usepackage{ulem}

\newcommand{\stkout}[1]{\ifmmode\text{\sout{\ensuremath{#1}}}\else\sout{#1}\fi}

\makeatother

\usepackage{babel}
\begin{document}
\title{Active Optical Frequency Measurements with  Superradiance Prolonged by a Modulated Magnetic Field}

\author{Huihui Yu}
\address{Henan Key Laboratory of Diamond Materials and Devices, Key Laboratory of Material Physics, Ministry of Education, School of Physics and Laboratory of Zhongyuan Light, Zhengzhou University, Zhengzhou 450052, China}

\author{Shi-Lei Su}
\address{Henan Key Laboratory of Diamond Materials and Devices, Key Laboratory of Material Physics, Ministry of Education, School of Physics and Laboratory of Zhongyuan Light, Zhengzhou University, Zhengzhou 450052, China}

\author{Chongxin Shan}
\email{cxshan@zzu.edu.cn}
\address{Henan Key Laboratory of Diamond Materials and Devices, Key Laboratory of Material Physics, Ministry of Education, School of Physics and Laboratory of Zhongyuan Light, Zhengzhou University, Zhengzhou 450052, China}

\author{Klaus M{\o}lmer}
\email{klaus.molmer@nbi.ku.dk}
\address{Niels Bohr Institute, University of Copenhagen, Blegdamsvej 17, 2100 Copenhagen, Denmark}

\author{Yuan Zhang}
\email{yzhuaudipc@zzu.edu.cn}
\address{Henan Key Laboratory of Diamond Materials and Devices, Key Laboratory of Material Physics, Ministry of Education, School of Physics and Laboratory of Zhongyuan Light, Zhengzhou University, Zhengzhou 450052, China}
\address{Institute of Quantum Materials and Physics, Henan Academy of Sciences, Zhengzhou 450046, China}

\begin{abstract}
Superradiant emission from long-lived excited states of an atomic ensemble confined in an optical cavity constitutes a practical source of light with narrow linewidth. In the pulsed regime, however, superradiance implies rapid emission and a broadening of the spectrum. Recent experiments have demonstrated constructive and destructive interference of superradiant emission by different strontium atomic transitions. In this article, we show that by modulating the atomic transition frequencies with a magnetic field, it is possible to control the release of the atomic excitation energy as a prolonged pulse or a train of superradiant pulses. By simulations, we show that heterodyne detection of the prolonged superradiance shows extremely sharp spectral features, which leads to significantly reduced frequency uncertainty and fluctuation.
\end{abstract}
\maketitle

\paragraph{Introduction}

Atomic clocks provide the ultimate precision time standard~\citep{BJaduszliwer}, underpinning both fundamental fields such as precision spectroscopy~\citep{JLHall}, tests of relativity~\citep{CHafeleJ}, and critical modern applications including GPS-based spacetime metrology and synchronization in high-speed communication~\citep{QShen}. For the atomic clocks, the transition from microwave ($\sim {\rm GHz}$) to optical frequencies ($\sim  10^5$ GHz) promises a reduction in Allan deviation by five orders of magnitude, targeting 
$10^{-18}$ for one second signal integration~\citep{LudlowAD, ABauch}. The pursuit of optical atomic clocks follows two paths with either individual trapped ions~\citep{ATofful} or with neutral atomic ensembles~\citep{MSchioppo}. An atomic ensemble trapped in an optical lattice suppresses the Doppler broadening effect, and permits a higher signal-to-noise ratio (SNR), approaching the theoretical value for the Allan deviation~\citep{TBothwell}.

\begin{figure}[!htp]
\begin{centering}
\includegraphics[scale=0.45]{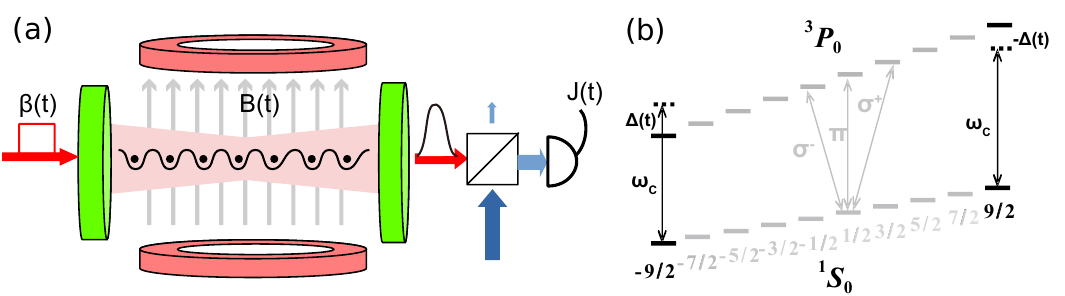}
\par\end{centering}
\caption{\label{fig:system} System schematic and energy level diagram. Panel (a) depicts optical lattice-trapped strontium-87 atoms, subject to a magnetic field $B\left(t\right)$, in an optical cavity field, driven by a laser pulse $\beta(t)$. The emitted signal is mixed with a reference laser beam and detected by a photodetector, generating the photocurrent  $J\left(t\right)$. Panel (b) shows the atomic energy diagram with ten hyperfine levels $m_{F}=-9/2,...,9/2$ (gray lines) of the electronic ground state $^{1}S_{0}$ and excited state $^{3}P_{0}$, and $\sigma^{+},\pi,\sigma^{-}$ transitions between these levels with $\Delta m_{F}=+1,0,-1$ (gray arrows). The black arrows show the nearly-resonant coupling of two atomic sub-ensembles in the extreme levels $m_{F}=-9/2,9/2$ to the cavity mode. The magnetic field gives rise to a time-dependent frequency detuning $\Delta(t)$.}
\end{figure}

While atom optical lattice clocks have demonstrated exceptional stability, they predominantly operate in a passive mode, locking a local oscillator to atomic transitions. In contrast, active optical clocks, where atoms function as a laser gain medium, offer a distinct paradigm characterized by wide detection bandwidth and high dynamic range~\citep{JChen2009,JChen2024}. The active atomic clock has been realized with hydrogen masers in the microwave domain~\citep{MAWeiss}, and recently, in the optical range with the strontium-based optical lattice clock~\citep{NorciaMA}. In the latter experiments, as illustrated in Fig. \ref{fig:system}, lattice-trapped strontium-87 atoms are excited by a laser pulse, and the excited atoms couple to an optical cavity with multiple oppositely detuned transitions.  The resulting superradiant emission is measured by heterodyne detection, and an Allan deviation of $6.7\times 10^{-16}/\sqrt{\tau/s}$  has been achieved as a function of the measurement time $\tau$ \citep{NorciaMA}. 

Control of superradiant beats by the magnetic field has been observed in experiments~\citep{NorciaMA2016}, and in this Letter, we apply theoretical methods~\citep{YZhang2021a} that combine quantum measurement theory~\citep{HWWiseman} and cavity quantum electrodynamics to analyze the system dynamics and unravel the physics underlying this novel phenomenon. Our simulations confirm that the superradiant beats are caused by constructive and destructive quantum interference between the atomic sub-ensembles. A modulated magnetic field can be applied to actively control the quantum interference by dynamically tuning the atomic transitions with respect to the cavity mode. We quantify how this mechanism allows us to actively prolong the superradiant pulses, and to observe a much sharper spectrum in the heterodyne detection. By exploring this spectrum, we further show that the frequency precision can be reduced by 2 and 7 times for one and ten seconds integration, and more importantly the frequency uncertainty  can be reduced by 26 times. Thus, our proposal constitutes an alternative solution for generating narrower coherent emission as compared with the superradiant laser~\citep{MANorcia,SLKristensen,HYu}. 

To describe the system shown in Fig.~\ref{fig:system}, we have developed a theory based on the stochastic master equation, and we apply a stochastic variant~\citep{HYu,ZQZhang} of the cumulant mean-field approach~\citep{DPlankensteiner} to account for the stochastic measurement backaction due to the heterodyne measurement [see Appendix A]. In the system, an optical cavity mode  with a frequency $\omega_c=2\pi \times 429.5$ THz and a damping rate $\kappa=2\pi\times 145$ kHz is  driven by a laser pulse of a frequency $\omega_d=\omega_c$ and strength $\beta(t)$ through the left mirror with the transmission coefficient $\sqrt{\kappa/2}$. More than  $N=2\times10^5$ strontium-87 atoms are evenly prepared in the two extreme sub-levels $m_{F}=-9/2,9/2$ of the singlet ground state $^{1}S_{0}$, and couple collectively to the cavity mode through their vertical $\pi$ transitions of frequencies $\omega_{m_F=-9/2,9/2}=\omega_{a}+\Delta_{B}m_{F}$ with the single-atom coupling strength $g=2\pi\times2.41$ Hz. Here, $\omega_{a}=\omega_c$ is the bare atomic transition frequency and \textbf{$\Delta_{B}=108.4\times B$} Hz is the proportionality between the $m_F$-dependent Zeeman shift and the applied magnetic field $B$ in Gauss.

\begin{figure}[!htp]
\begin{centering}
\includegraphics[scale=0.13]{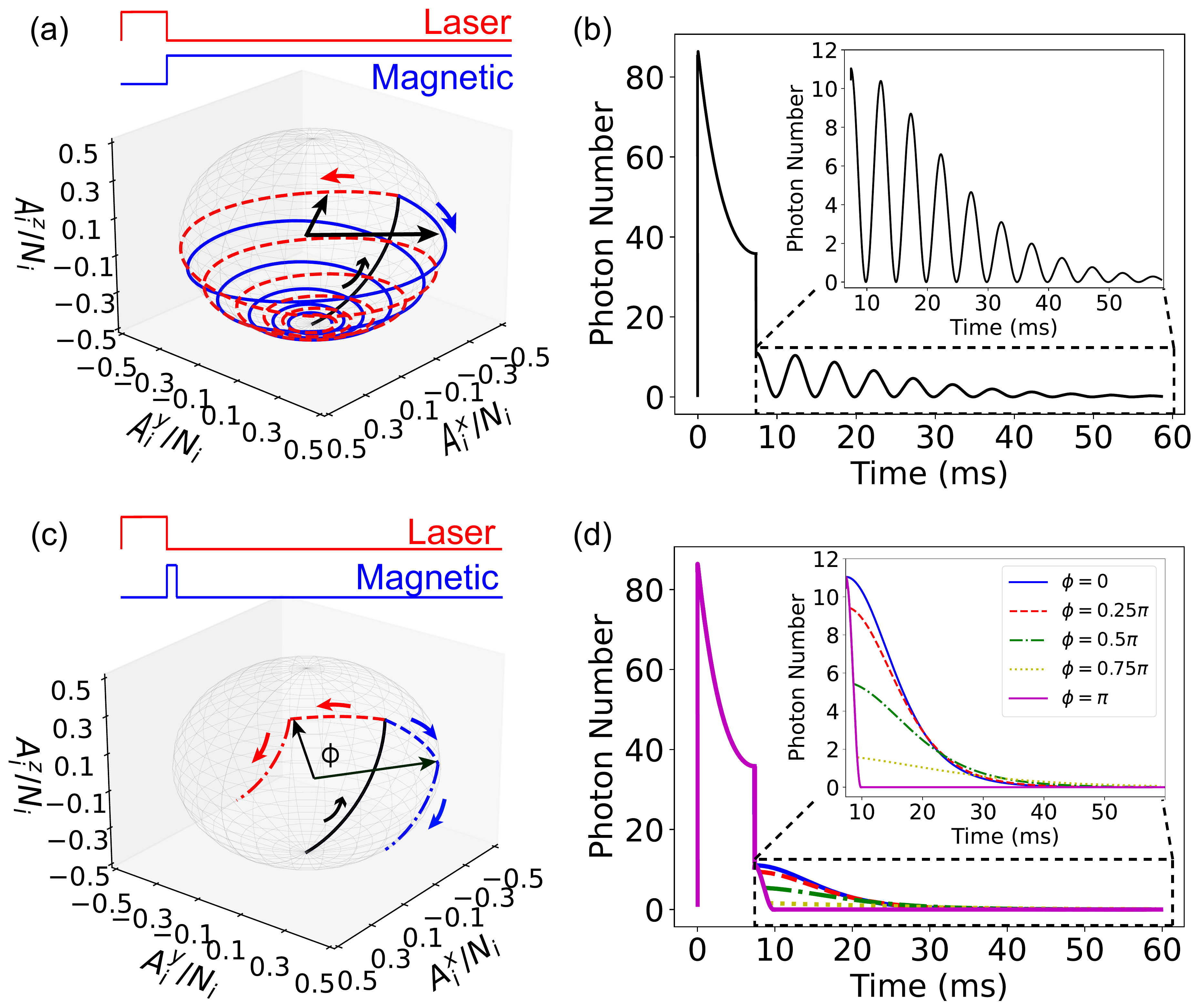}
\par\end{centering}
\caption{\label{fig:beats} Magnetic field-controlled superradiance. Panel (a) shows the sequential application of a laser pulse and a magnetic field (top), causing the rotation and subsequent precession of the Bloch vectors ${\bf A}_{9/2}$,${\bf A}_{-9/2}$ of the two atomic sub-ensembles (red dashed and blue solid curves). Panel (b) shows how the intra-cavity photon number decays after its transient excitation by the input pulse and then follows the oscillating coherent atomic dipole moment resulting from addition of the two Bloch vectors. Panel (c) illustrates the rotations of the two Bloch vectors subjected to the laser and magnetic field pulses shown at the top of the panel. Panel (d) shows the monotonous decay of the intra cavity photon number when the atomic precession halts. The insets in panels (b,d) show a zoom-in of the intra-cavity photon number for longer time. The magnetic field controls the angle $\phi$, which is equal to $\pi/2$ in panel (c), and varies from zero to $\pi$ to control the photon number in panel (d). } 
\end{figure}

\paragraph{Magnetic Field-controlled Superradiance} The Zeeman effect causes atomic sub-ensembles to evolve coherently at different frequencies and lead to superradiant beats~\citep{NorciaMA2016,NorciaMA}. This behavior is well captured by two Bloch vectors ${\bf A}_{9/2}$ and ${\bf A}_{-9/2}$, representing the quantum state of the atoms in the respective Zeeman sub-ensembles. These vectors point to the  south and north pole for the atoms in the ground and excited states, respectively, while other directions represent the superposition states. To observe the superradiant beats in our simulations, we first drive the optical cavity with a laser pulse, and then introduce a magnetic field, which causes the two Bloch vectors to rotate clockwise and anti-clockwise while they gradually descend towards the ground state [Fig. \ref{fig:beats}(a)]. The intra-cavity photon number forms superradiant beats with maxima when the horizontal projections of the two Bloch vectors are parallel and minima when they are anti-parallel [Fig. \ref{fig:beats}(b)]. The superradiant beats are thus clearly caused by constructive and destructive interference of the two atomic sub-ensembles. Note that here the oscillations are not vacuum Rabi oscillations since the system works in the bad-cavity regime. 

Based on this simple picture, we note that the dynamics can in principle be halted at minimum emission if we can establish and maintain anti-parallel directions of the horizontal components of the Bloch vectors by application of a suitable time-dependent magnetic field. We may thus suppress the precession, stretch the duration of the  superradiant emission or switch it on and  off at chosen times. Figure \ref{fig:beats}(c,d) thus show that after being excited to the equator plane, the two Bloch vectors rotate clockwise and anti-clockwise only when the magnetic field is on, forming an angle $\phi$, and hereafter they rotate around the y-axis and converge to points on the Bloch sphere  with anti-parallel horizontal projections. After the superradiant pulse, the two atomic sub-ensembles become decoupled from the cavity, and the Bloch vectors become effectively frozen in the short time scale about tens of microseconds. However, in a longer time scale about hundreds of seconds, the atoms can still decay gradually to the ground state through the spontaneous emission, leading to the further rotation of the Bloch vectors to the south pole. Note that here the rotation axis is controlled by the phase of the atoms-cavity coupling $g$. By changing the length of the magnetic pulse, we can actively control the initial angle $\phi$, and thus the subsequent atomic decay and photoemission dynamics [see the inset of Fig. \ref{fig:beats}(d)].

\begin{figure}[!htp]
\begin{centering}
\includegraphics[scale=0.13]{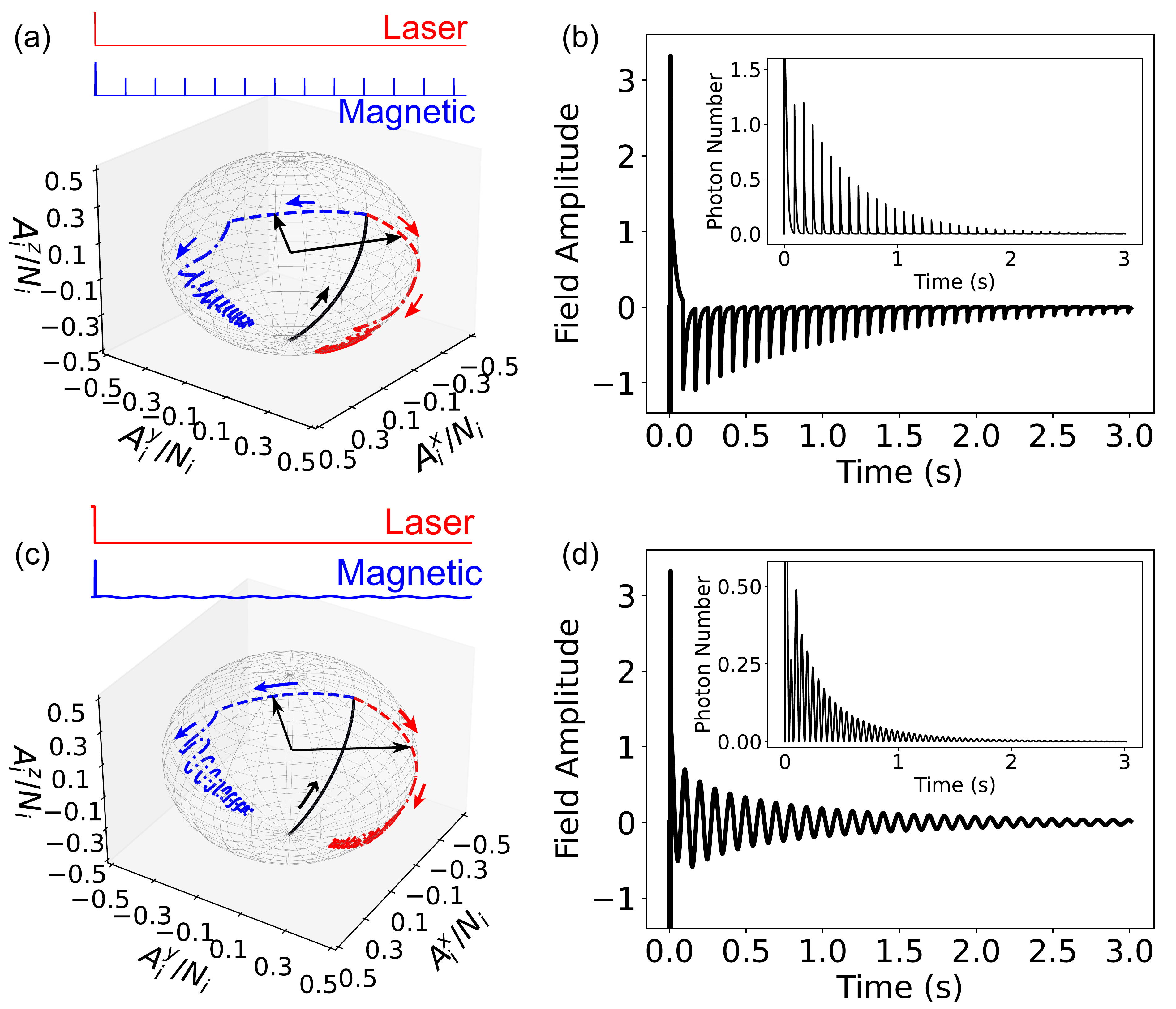}
\par\end{centering}
\caption{\label{fig:pulses} Superradiance engineered by a train of short magnetic pulses (a,b) and by a sinusoidal modulation of the magnetic field (c,d). The corresponding frequency detuning $\Delta(t)$ has an amplitude of $2\pi\times100$ Hz and a period of $0.63$ ms for the pulses shown at the top of panel (a), and an amplitude $2\pi\times2.5$ Hz and a period $0.1$ s for the modulation shwon at the top of panel (b). Panels (a,c) show the dynamics of the sub-ensemble Bloch vectors and panels (b,d) show the dynamics of the intra-cavity field amplitude and photon number (insets).}
\end{figure}


The above calculations show that, independent of the angle $\phi$, during the atomic decay the Bloch vectors of the two atomic sub-ensembles develop anti-parallel horizontal projections and the emission halts. By applying yet another magnetic pulse, we can cause the Bloch vectors to depart from the anti-parallel projections so that a net atomic coherence appears and superradiant emission continues. By repeating the magnetic pulses, we may thus expect to generate a train of superradiant pulses and this is, indeed, confirmed by our numerical analysis, as shown in Fig. \ref{fig:pulses} (a,b). Rather than multiple square magnetic pulses, it may be easier to generate a smooth sinusoidally varying magnetic field, and simulations for this situation are shown in Fig. \ref{fig:pulses}(c,d). We see that the two Bloch vectors oscillate around a trajectory leading to the anti-parallel horizontal projections [Fig. \ref{fig:pulses}(c)], and the resulting superradiance shows smooth oscillations with a gradually reduced amplitude [Fig. \ref{fig:pulses}(d)]. 
Since the Bloch vectors are perturbed to the same side and are forced to oscillate for the case with the square and sinusoidal magnetic fields, the field amplitude is always negative in the former case but changes the sign periodically in the latter case. In both cases, the duration of superradiance is about $30$ times longer than that without the magnetic field modulation [inset of Fig. \ref{fig:pulses}(b,d) and Fig.~\ref{fig:beats}(b,d)]. In Fig. A2 in the Appendix C, we illustrate how the amplitude of the frequency oscillations can be used to further control the amplitude and length of superradiant pulses.

\begin{figure}[!htp]
\begin{centering}
\includegraphics[scale=0.24]{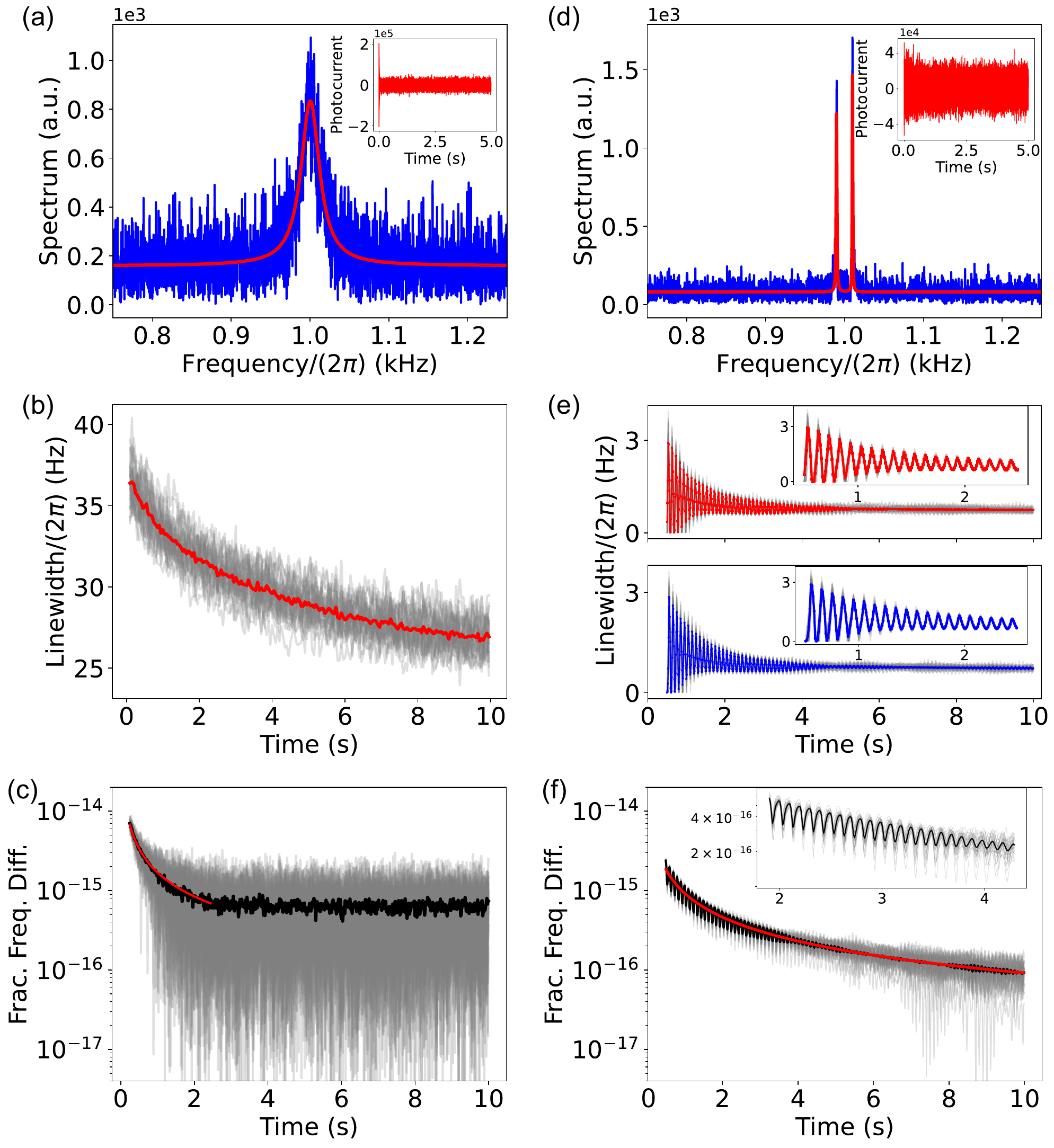}

\par\end{centering}
\caption{\label{fig:measurement} Active optical frequency measurements with short (a,b,c) and prolonged (d,e,f) superradiant pulses as shown in Fig. \ref{fig:beats} (d) and Fig. \ref{fig:pulses}(d). Panels (a) and (d) show the power spectra (blue), obtained by Fourier transform of $\tau=5$ s long simulated  noisy photocurrents (insets), and corresponding Lorentzian fits (red curves). Panels (b) and (e) show the dependence of the extracted linewidth on the duration of the measurement, where the colored and gray curves show the mean and the overlap of 30 simulations respectively. In panel (e), the upper and lower part show the results for the left and right frequency peaks shown in panel (d). Panels (c) and (f) show the mean (black curves) of the extracted fractional frequency difference over many simulations (overlapped gray curves), as a function of the duration of the measurement with the short and long superradiant pulses. The means are fitted with the expressions shown in the text. The insets of panels (e) and (f) display the magnified views of the local regions.}
\end{figure}

\paragraph{Improved Frequency Measurement with Prolonged Superradiance }

The above simulations demonstrate that the superradiant pulse can be prolonged in time by modulating the magnetic field. In the following, we investigate whether this prolonged signal can improve the precision of the optical frequency measurement. To this end, we consider heterodyne detection, where the signal is mixed with a local oscillator of frequency $\omega_l=\omega_c + 2\pi$ kHz, and the mixed beam is measured by a photodetector of efficiency $\eta=0.05$. We can simulate the noisy photocurrent obtained in an experiment with the expression, $J\left(t\right)=\sqrt{2\eta\kappa}\mathrm{Re}\left[e^{-i\omega_{l}t}\left\langle \hat{a}^{\dagger}\right\rangle \right]+dW/dt$, which has a  linear deterministic term, proportional to the cavity field amplitude $\left\langle \hat{a}^{\dagger}\right\rangle$, and a photon-shot-noise contribution $dW/dt$, dominating the photocurrent in each short time interval $dt$. The mean field amplitude can be calculated with the cumulant expansion method, and $dW$ is simulated as a normally distributed random number with a vanishing mean $E[dW\left(t\right)]=0$ and a variance $dW\left(t\right)^{2}=dt$. The backaction of the measurement on the system is represented by an accompanying conditional master equation~\citep{HWWiseman}, where the stochastic signal influences the quantum state and hence the subsequent spectral correlations in the measurement signal~\citep{YZhang2021a,HYu}.

As a reference, we show in Fig. \ref{fig:measurement}(a) the power spectrum of the heterodyne signal,  simulated for the situation studied in the experiments~\citep{NorciaMA}. The photocurrent shows large fluctuations in the first $50$ ms and converges to smaller steady state fluctuations for longer  times [inset of Fig.~\ref{fig:measurement}(a)], where the envelope resembles the calculated superradiant signal [Fig. \ref{fig:beats}(d)]. The Fourier transform of the noisy signal for a duration of 5 s leads to a power spectrum with a single peak at around $2\pi$ kHz over a noise background, i.e., at the detuning of the local oscillator from the cavity mode frequency. By fitting this peak with a Lorentzian function, we obtain a center frequency $ \omega/2\pi = 1000.09$ Hz and a linewidth  $\delta = 29.03$ Hz. Note that the extracted frequency is only $0.09$ Hz away from the expected value.

In Fig.~\ref{fig:measurement}(d), we show similar results for the case where the superradiant signals are prolonged over time by the magnetic field control. In the case with a sinusoidally varying magnetic field, the power spectrum shows two extremely sharp peaks on a nearly negligible background. By fitting these peaks with Lorentzian functions, we obtain the frequencies  $\omega_{1}/(2\pi)=989.98$ Hz, $\omega_{2}/(2\pi)=1010.18$ Hz and linewidths $\delta_{1}/(2\pi)=0.79$ Hz, $\delta_{2}/(2\pi)=0.75$ Hz. Their deviation from $(\omega_l-\omega_a)/(2\pi)$ is $10$ Hz, which is the frequency of the sinusoidal magnetic field pulse $\omega_m$. Owing to the substantially longer pulse duration, these linewidths are close to the Fourier limit and are approximately $37$ times narrower than those of the short pulse case without magnetic field control. We have also simulated the case with square magnetic field pulses and obtained noisy photocurrents similar to the ones shown in Fig. \ref{fig:pulses}(b). However, since the corresponding power spectrum shows several sharp spikes over a broad peak (Fig. A3 in Appendix C), it is more difficult to extract well-defined frequencies.  

To quantify the dependence of the power spectrum on the measurement,  we calculate the Fourier transform of the photocurrent up to a variable finite measurement time $\tau$, and fit the obtained spectra with a Lorentzian function for each peak for many simulations, and obtain a collection of fitted linewidths $\delta(\tau)$ and center frequencies $\omega(\tau)$. In the case with the short pulse [Fig. \ref{fig:measurement}(b)], as the measurement time increases, the extracted linewidth $\delta$ decreases relatively slowly from about $37$ Hz to $27$ Hz, and the variation among different simulations remains constant. The linewidth is continuously improved for longer measurement time because the frequency resolution is increased for longer sampling time. However, since the detection samples only noise after the pulse has expired, the SNR of the peaks decreases gradually for longer measurement time (Fig. A4 in Appendix C). In the case with the prolonged superradiant pulses [Fig. \ref{fig:measurement}(e)], the linewidth oscillates with a reduced amplitude with the same period as the magnetic field modulation, and it gradually approaches  $0.73$ Hz, which is about $36$ times smaller than that of the short pulse. More importantly, the SNR of the peaks decays slowly after increasing for short measurement times, but it is   larger than that of the short pulse (Fig. A4 in Appendix C), which we attribute to the persistence of the superradiant signal during the measurement. 
 
To quantify the precision of frequency measurement, we calculate the fractional frequency difference (FFD), as a function of the measurement time $\tau$, see Fig. \ref{fig:measurement} (c) and (f). In the case with the short pulse, we calculate the FFD with the expression $\sigma(\tau) = |\omega(\tau)-\omega_l|/\omega_a$, where $\omega_l,\omega_a$ are the frequencies of the local oscillator and the bare atomic transition, and $\omega(\tau)$ is the extracted frequency for given measurement length $\tau$. We see that the mean FFD scales with the expression $1.69\times10^{-15}/(\tau/s)$ for the short $\tau$, and reaches the saturated value around $6.33\times 10^{-16}$ for $\tau$ larger than $4$ s. These results are characteristic of active frequency measurements~\citep{MAWeiss} and they agree qualitatively with the experiment~\citep{NorciaMA} and with the theoretical results~\citep{YZhang2021a}. In addition, the fluctuations of the FFD among different simulations (gray curves) seem to become comparable or even larger than the mean of FFD with increasing $\tau$. To analyze this in more detail, we have plotted the standard deviation of the FFD as a function of $\tau$, and found that it remains at the level around $7.69\times 10^{-16}$ [Fig. A4 in Appendix C]. In the case of the prolonged pulse, we define the FFD with the expressions $|[\omega_1(\tau)+\omega_2(\tau)]/2-\omega_l|/\omega_a$, where $\omega_1(\tau),\omega_2(\tau)$ are the extracted frequencies of the left and right peaks shown in Fig.~\ref {fig:measurement}(d). We find that the FFD exhibits oscillations in analogy  to those of the linewidth, and the average FFD and the oscillation amplitude gradually decay as the measurement time $\tau$ increases. The averaged FFD can be fitted as $9.27\times10^{-16}/(\tau/s)$, and decreases to $9.27\times 10^{-17}$ at $\tau =10$ s, which is $7$ times smaller than that achieved with the short pulse [Fig.~\ref {fig:measurement}(c)]. The fluctuations of the FFD among different simulations are much smaller (gray curves), and the associated standard deviation is about 26 times smaller than for the short pulse (Fig. A4 in Appendix C).
In addition, we have also verified that the decay and decoherence of the optical clock transition have negligible influence on the frequency measurement based on the superradiance (Fig. A5 in Appendix C).

\paragraph{Conclusion}
In summary, we have proposed and theoretically validated the possibility to prolong superradiant pulses from strontium atomic ensembles  by controlling the constructive and destructive quantum interference of two sub-ensembles with a modulated magnetic field. We have shown that the heterodyne detection of a single superradiant pulse leads to a single broad peak, while detection of the prolonged pulse displays two narrower peaks on a nearly negligible background. By exploring these narrower peaks, our simulations show that the frequency precision can be reduced by 2 and 7 times for one- and ten-second integration time, and the frequency fluctuations can be reduced by a factor of 26. We note that steady-state superradiance~\citep{DMeiser2009,SLKristensen2023} constitutes a promising ultra-narrow frequency light source, while its continuous pumping may be a technical challenge and a source of phase noise. A quantitative comparison should be made between the performance of prolonged pulses and steady-state superradiance, and such analyses may also apply to other optical clock systems~\citep{TLaske}.

\begin{acknowledgments}
The authors thank James K. Thompson for providing helpful insights into the related superradiance experiments. Huihui Yu and Shi-Lei Su contribute equally to this work. This work was supported by the Scientific Research Innovation Capability Support Project for Young Faculty No. SRICSPYF-BS2025008, and the National Natural Science Foundation of China through the project No. 12422413 and No. 62475242, Beijing National Laboratory for Condensed Matter Physics No. 2023BNLCMPKFO17, Zhengzhou University Young Student Basic Research Projects (PhD students) No. ZDBJ2026043, as well as the Carlsberg Foundation through the "Semper Ardens" Research Project QCooL.
\end{acknowledgments}

\appendix
\renewcommand\thefigure{A\arabic{figure}}
\renewcommand\thetable{A\arabic{table}}
\renewcommand{\theequation}{A\arabic{equation}}
\setcounter{figure}{0}
\setcounter{equation}{0}

\makeatletter
\@removefromreset{equation}{section}  
\makeatother
\section{Stochastic Master Equation for Heterodyne Detection \label{sec:sme}}
In our previous article~\citep{YZhang2021a,HYu}, we have proposed a stochastic master equation for the heterodyne detection of the superradiance. We apply the modified QuantumCumulants.jl package~\citep{DPlankensteiner} to solve the stochastic master equation for the density operator $\hat{\rho}_{J}$ conditioned on the photocurrent $J(t)$ of the heterodyne detection:
\begin{align}
 & \partial_t\hat{\rho}_{J}=-\frac{i}{\hbar}\left[\hat{H}_{c}+\hat{H}_{d}+\hat{H}_{a}+\hat{H}_{a-c},\hat{\rho}_{J}\right]\nonumber \\
 & -\frac{\kappa}{2}\left(\hat{a}^{\dagger}\hat{a}\hat{\rho}_{J}+\hat{\rho}_{J}\hat{a}^{\dagger}\hat{a}-2\hat{a}\hat{\rho}_{J}\hat{a}^{\dagger}\right)\nonumber \\
 & +\frac{dW}{dt}\sqrt{\eta\kappa/2}\bigl[e^{i\omega_{l}t}\left(\hat{a}-\left\langle \hat{a}\right\rangle \right)\hat{\rho}_{J}+h.c.\bigr].\label{eq:cme}
\end{align}
The first
line of Eq. (\ref{eq:cme}) describes the coherent dynamics. The cavity
mode Hamiltonian $\hat{H}_{c}=\hbar\omega_{c}\hat{a}^{\dagger}\hat{a}$
is specified with a frequency $\omega_{c}$, a photon creation $\hat{a}^{\dagger}$
and annihilation operator $\hat{a}$. The driving of the cavity mode
by a laser $\hat{H}_{d}=\sqrt{\kappa/2}\hbar\beta\left(t\right)e^{i\omega_{d}t}\hat{a}+h.c.$
is given by the laser strength $\beta\left(t\right)$, frequency
$\omega_{d}$ and transmission coefficient $\sqrt{\kappa/2}$ through the left mirror. The atomic Hamiltonian $\hat{H}_{a}=\hbar\sum_{i=9/2,-9/2}\omega_{i}\sum_{k=1}^{N_{i}}\hat{\sigma}_{i,k}^{22}$ describes the two sub-ensembles (specified by $i=m_{F}=-F,+F$ with $F=9/2$) of   $N_{i}$ atoms (labeled by $k$), the frequencies $\omega_{i=9/2,-9/2}=\omega_{a}+\Delta_{B}m_{F}$,
which depend on the bare frequency $\omega_{a}/2\pi=429.5$ THz and the Zeeman shift \textbf{$\Delta_{B}=108.4\times B$} Hz for a static magnetic field $B$ in Gauss, as well as the upper-state projection operator $\hat{\sigma}_{i,k}^{22}$ of individual atoms. The atom-cavity mode interaction Hamiltonian $\hat{H}_{a-c}=\hbar\sum_{i=9/2,-9/2}g_{i}\left[\hat{a}^{\dagger}\left(\sum_{k}\hat{\sigma}_{i,k}^{12}\right)+h.c.\right]$
depends on the coupling strengths $g_{i=m_{F}}=g_{0}m_{F}/\sqrt{F\left(F+1\right)}$,
determined by a constant $g_{0}=2\pi\times2.41$ Hz and Clebsch-Gordan coefficients, and the lowering $\hat{\sigma}_{i,k}^{12}$ (and raising
$\hat{\sigma}_{i,k}^{21}$) ladder operators of the individual atoms. 

The second line of Eq. (\ref{eq:cme}) describes the photon loss with a rate $\kappa=2\pi \times 145 $ kHz due to the left ($\kappa/2$) and right ($\kappa/2$) mirrors. Here, we ignore the negligible spontaneous emission and dephasing of the optical lattice clock system. The third line of Eq. (\ref{eq:cme})
accounts for the measurement backaction~\citep{HWWiseman} with the detector photon shot noise $dW$, and the photon-counting efficiency $\eta$ of the detector, as well as the frequency $\omega_{l}$ of the local oscillator. The photon shot-noise follows a normal distribution with a variance $dW\left(t\right)^{2}=dt$ and a mean  $E\left[dW\left(t\right)\right]=0$, where $dt$ is the time-step in the numerical simulations. 

To simulate the system with tens of thousands of atoms, we cannot solve the conditional master equation (\ref{eq:cme}) 
with the standard density matrix technique because of the exponentially increased Hilbert space with the number of atoms, and thus adopt here the cumulant mean-field approach~\citep{KDebnath,YZhang2021a}. In the mean-field approach, we
derive the equation $\partial_t\left\langle \hat{o}\right\rangle =\mathrm{tr}\left\{ \left(\partial_t\hat{\rho}\right)\hat{o}\right\} $
for the mean value $\left\langle \hat{o}\right\rangle =\mathrm{tr}\left\{ \hat{\rho}\hat{o}\right\} $
of any operator $\hat{o}$, and truncate the resulting hierarchy of equations by applying the cumulant expansion approximation, and obtain a closed set of equations for mean-field quantities of given order. Assuming that the atoms in one sub-ensemble have the same transition frequencies and coupling with the cavity mode, as considered above, we can explore the symmetry inherited in the mean fields to reduce dramatically the number of independent equations. The mean-field approach has been implemented in the QuantumCumulants.jl package~\citep{DPlankensteiner} to solve the deterministic master
equation, and in the present article, we generalize the implementation to the stochastic master equation. 
We present the Julia codes to derive and solve the mean-field equations in Appendix. \ref{sec:JuliaCodes}.
We obtain the mean-field equations to first order with the approximation $\langle\hat{o}\hat{p}\rangle \approx \langle\hat{o}\rangle\langle\hat{p}\rangle$ (for any operator $\hat{o},\hat{p}$). The field amplitude $\langle\hat{a}^{\dagger}\rangle$ is defined as the expectation value ${\rm tr} \{\hat{a}^{\dagger} \hat{\rho}_J\}$ of the photon creation operator $\hat{a}^{\dagger}$, and satisfies the equation 
\begin{align}
 & \partial_{t}\langle\hat{a}^{\dagger}\rangle=i\left(\omega_{c}+i\kappa/2\right)\langle\hat{a}^{\dagger}\rangle+i\beta(t)\sqrt{\kappa/2}e^{i\omega_{d}t}\nonumber \\
 & +iN_{1}g_{1}\langle\hat{\sigma}_{1,1}^{21}\rangle+iN_{2}g_{2}\langle\hat{\sigma}_{2,1}^{21}\rangle. \label{eq:ap1st}
\end{align}
This equation couples to the atomic coherence $\langle\hat{\sigma}_{1,1}^{21}\rangle,\langle\hat{\sigma}_{2,1}^{21}\rangle$
in the two atomic sub-ensembles. Here and in the following, for the operators $\hat{\sigma}_{i,j}^{mn}$, the sub-index $i=1,2$ indicates the atomic sub-ensembles with $m_F=-9/2,9/2$, and $j=1,2$ labels the first and second representative atom for given sub-ensemble, and the upper-indices $m,n=1,2$ label the atomic ground and excited state. Thus, the operators $\hat{\sigma}_{i,j}^{11},\hat{\sigma}_{i,j}^{22}$ are the projection operators, while $\hat{\sigma}_{i,j}^{12},\hat{\sigma}_{i,j}^{21}$ are the lowering and raising ladder operators. The atomic coherences $ \langle\hat{\sigma}_{i,1}^{21}\rangle$ for $i=1,2$ satisfy
the equations 
\begin{equation}
\partial_{t} \langle\hat{\sigma}_{i,1}^{21}\rangle=i\omega_{i}\langle\hat{\sigma}_{i,1}^{21}\rangle-ig_{i}\langle a^{\dagger}\rangle\left(2\langle\hat{\sigma}_{i,1}^{22}\rangle-1\right).\label{eq:sig221}
\end{equation}
These equations depend on the populations $\langle\hat{\sigma}_{1,1}^{22}\rangle,\langle\hat{\sigma}_{2,1}^{22}\rangle$ of the  excited states for the atoms in the first and second sub-ensemble, respectively. The populations $\langle\hat{\sigma}_{i,1}^{22}\rangle$ for $i=1,2$ satisfy the equations
\begin{equation}
\partial_{t}\langle\hat{\sigma}_{i,1}^{22}\rangle  =ig_{i}\left(\langle\hat{a}^{\dagger}\rangle\langle\hat{\sigma}_{i,1}^{12}\rangle-\langle a\rangle\langle\hat{\sigma}_{i,1}^{21}\rangle\right). \label{eq:sig222}
\end{equation}
Note that $\langle a\rangle$ is the expectation value  ${\rm tr} \{\hat{a} \hat{\rho}_J\}$ of the photon annihilation operator $\hat{a}$, and is the complex conjugate of $\langle\hat{a}^{\dagger}\rangle$. To visualize the atomic dynamics, we introduce the normalized Bloch
vectors $\textstyle
\mathbf{A}_{i}/N_{i}=\mathrm{Re}\langle \hat{\sigma}_{i,1}^{21}\rangle \mathbf{e}_{x}+\mathrm{Im}\langle \hat{\sigma}_{i,1}^{21}\rangle \mathbf{e}_{y}+(2\langle \hat{\sigma}_{i,1}^{22}\rangle -1)\mathbf{e}_{z}
$
in the Bloch sphere, where $\mathbf{e}_{k=x,y,z}$ are the unit vectors of the Cartesian coordinate system. 
To obtain the mean-field equations to second order, 
we apply the approximation $\langle\hat{o}\hat{p}\hat{q}\rangle \approx \langle\hat{o}\rangle\langle\hat{p}\hat{q}\rangle + \langle\hat{p}\rangle\langle\hat{o}\hat{q}\rangle + \langle\hat{q}\rangle\langle\hat{o}\hat{p}\rangle -2\langle\hat{o}\rangle\langle\hat{p}\rangle\langle\hat{q}\rangle$ (for any operator $\hat{o},\hat{p},\hat{q}$), which are obtained from vanishing third-order cumulant.
In these equations, we encounter not only the first-order quantities, but also second-order quantities, e.g. the mean photon number $\langle\hat{a}^{\dagger}\hat{a}\rangle$,
the photon-photon correlation $\langle\hat{a}\hat{a}\rangle$, the
atom-photon correlations $\langle\hat{a}^{\dagger}\hat{\sigma}_{i,1}^{12}\rangle,\langle\hat{a}^{\dagger}\hat{\sigma}_{i,1}^{22}\rangle,\langle\hat{a}\hat{\sigma}_{i,1}^{12}\rangle$
(with $i=1,2$), the correlations $\langle\hat{\sigma}_{i,1}^{21}\hat{\sigma}_{j,2}^{12}\rangle,\langle\hat{\sigma}_{i,1}^{22}\hat{\sigma}_{i,2}^{21}\rangle,\langle\hat{\sigma}_{i,1}^{12}\hat{\sigma}_{i,2}^{12}\rangle,\langle\hat{\sigma}_{i,1}^{22}\hat{\sigma}_{i,2}^{22}\rangle$
between the atoms of the same sub-ensemble ($i=1,2$), the correlations
$\langle\hat{\sigma}_{1,1}^{12}\hat{\sigma}_{2,1}^{21}\rangle,\langle\hat{\sigma}_{1,1}^{22}\hat{\sigma}_{2,1}^{21}\rangle,\langle\hat{\sigma}_{1,1}^{21}\hat{\sigma}_{2,1}^{22}\rangle,\langle\hat{\sigma}_{1,1}^{12}\hat{\sigma}_{2,1}^{12}\rangle,\langle\hat{\sigma}_{1,1}^{22}\hat{\sigma}_{2,1}^{22}\rangle$
between the atoms of the two sub-ensembles. As an example, here, we show the equation for the field amplitude 
\begin{align}
 & \partial_{t}\langle\hat{a}^{\dagger}\rangle=i\left(\omega_{c}+i\kappa/2\right)\langle\hat{a}^{\dagger}\rangle+i\sqrt{\kappa/2}\beta(t) e^{i\omega_{d}t}\nonumber \\
 & +iN_{1}g_{1}\langle\hat{\sigma}_{1,1}^{21}\rangle+iN_{2}g_{2}\langle\hat{\sigma}_{2,1}^{21}\rangle\nonumber \\
 & + \frac{dW}{dt} \sqrt{\xi\kappa/2}e^{i\omega_{l}t}\left(\langle\hat{a}^{\dagger}\hat{a}\rangle-\langle\hat{a}^{\dagger}\rangle\langle\hat{a}\rangle\right)\nonumber \\
 & + \frac{dW}{dt} \sqrt{\xi\kappa/2}e^{-i\omega_{l}t}\left(\langle\hat{a}^{\dagger}\hat{a}^{\dagger}\rangle-\langle\hat{a}^{\dagger}\rangle^{2}\right).\label{eq:ap2nd}
\end{align}
A comparison of Eqs. (\ref{eq:ap1st}) and (\ref{eq:ap2nd}) indicates that the measurement does not affect the dynamics and adds merely noise to the photocurrent in the first-order mean-field equations. However, it affects both the dynamics and the photon-current in the second-order mean-field equations. Note that all the results shown in the main text are calculated with the second-order mean-field equations. 
The results shown in Fig. 2(b,d) in the main text can be obtained through both the first- and second-order mean-field equations, and the measurement backaction does not affect the system dynamics for the current situation. However, we emphasize that the measurement backaction as described correctly by the second-order mean-field equations is essential to account for the heterodyne detection of superradiance~\citep{HYu} from the atomic ensembles initially prepared in the full excited state~\citep{YZhang2021a} or incoherently excited from the ground state~\citep{KDebnath,YZhang,DMeiser2009}.

\begin{figure}
\begin{centering}
\includegraphics[scale=0.29]{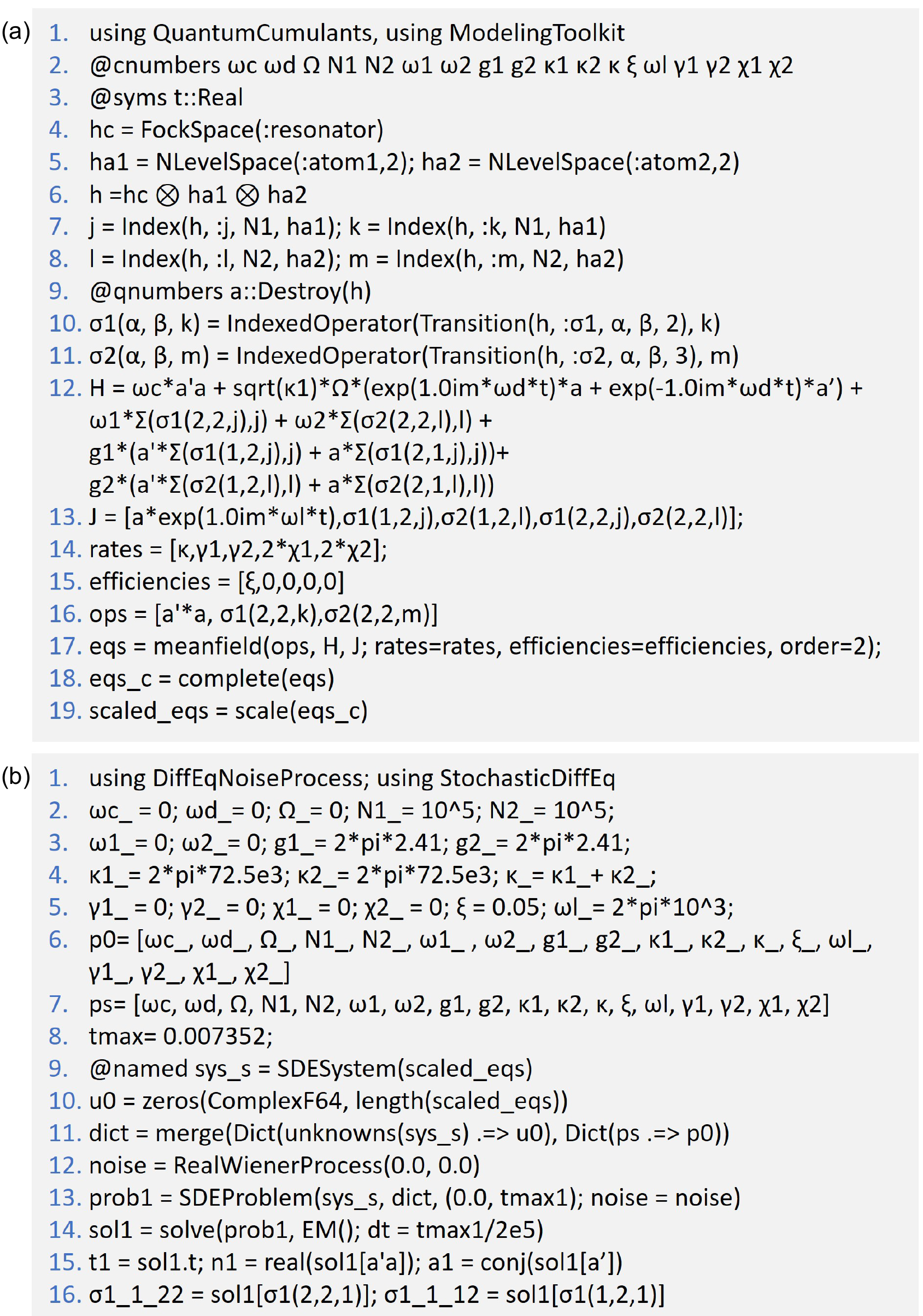}

\par\end{centering}
\caption{\label{fig:JuliaCode}Julia code for deriving and solving stochastic equations. Panel (a) provides the code to derive symbolic second-order mean-field equations for the relevant quantities. Panel (b) displays the code to solve the resulting numerical stochastic equations. First-order mean-field equations can be obtained by setting the parameter "order" to 1.}
\end{figure}

\section{Julia Codes to Solve Stochastic Master Equation \label{sec:JuliaCodes}}

In this section, we describe the Julia code to derive the equations for the mean-field quantities to second order from the stochastic master equation [see Fig. \ref{fig:JuliaCode}(a)], and the code to solve the numerical equations with stochastic differential equation method [see Fig. \ref{fig:JuliaCode}(b)].

Lines 1 through 3 in Fig. \ref{fig:JuliaCode} (a) import the QuantumCumulants.jl and ModelingToolkit.jl packages, define the symbolic complex parameters, and define the real time variable $t$. Lines 4 through 6 construct the composite Hilbert space: line 4 defines the Fock space for the cavity mode, line 5 defines the Hilbert spaces for the two sub-ensembles of two-level atoms, and line 6 forms the total Hilbert space for the atom-cavity system via the tensor product. Lines 7 and 8 define the indices used to label different atoms within the two sub-ensembles. Line 9 defines the photon annihilation operator $a=\hat{a}$. Lines 10 and 11 define the transition operators $\sigma1(\alpha,\beta,k)$ and $\sigma2(\alpha,\beta,m)$ for the two sub-ensembles, respectively. Line 12 defines the system Hamiltonian with $\hbar=1$ and $a'=\hat{a}^{\dagger}$. Lines 13 and 14 specify the list of jump operators and their corresponding rates to construct the Lindblad superoperators. Line 15 defines the measurement efficiencies. Lines 16 and 17 specify the list of operators, including the photon number operator $a'a=\hat{a}^{\dagger}\hat{a}$ and the upper-state projection operators $\sigma1(2,2)[k]=\hat{\sigma}_{1,k}^{22},\sigma2(2,2)[m]=\hat{\sigma}_{2,m}^{22}$ (with $k=1$), and derive the equations for the expectation values of any atom in the first and second ensembles. Line 18 uses the complete function to close the system. Line 19 uses the scale function to scale the equations for numerical evaluation by treating all atoms within each ensemble as identical.

During the derivation of equations, we carry out different orders
of approximation. For the first-order mean-field approach, we assume the negligible second-order cumulant $\left\langle \hat{p}\hat{q}\right\rangle _{c}=\left\langle \hat{p}\hat{q}\right\rangle -\left\langle \hat{p}\right\rangle \left\langle \hat{q}\right\rangle \approx0$
to approximate the mean-values in second order $\left\langle \hat{p}\hat{q}\right\rangle $ by the product $\left\langle \hat{p}\right\rangle \left\langle \hat{q}\right\rangle $. For the second-order mean-field approach, we assume the negligible third-order cumulant to approximate the mean-values in third order $\left\langle \hat{o}\hat{p}\hat{q}\right\rangle $
by the products $\left\langle \hat{o}\right\rangle \left\langle \hat{p}\hat{q}\right\rangle +\left\langle \hat{o}\hat{q}\right\rangle \left\langle \hat{p}\right\rangle +\left\langle \hat{q}\right\rangle \left\langle \hat{o}\hat{p}\right\rangle -2\left\langle \hat{o}\right\rangle \left\langle \hat{p}\right\rangle \left\langle \hat{q}\right\rangle $. Here, $\hat{o},\hat{p},\hat{q}$
stand for any operators.

Figure \ref{fig:JuliaCode}(b) shows the Julia code to define the system of the stochastic differential equations. Line 1 imports the DiffEqNoiseProcess.jl and StochasticDiffEq.jl packages. Lines 2 through 8 specify the values for the involved parameters, construct a list of values and parameters, and set the maximum simulation time. Line 9 constructs the symbolic SDE system from previously derived scaled mean-field equations using SDESystem. Line 10 initializes the mean-field variables to zero. Line 11 creates a dictionary mapping the symbolic unknowns and parameters to their numerical values and initial conditions. Line 12 defines a real Wiener process to represent the stochastic measurement noise. Lines 13 and 14 construct the SDE problem and solve the SDE using the Euler-Maruyama method. Lines 15 and 16 extract from the simulation the list of simulation time, the mean photon number, the upper-state population, and the atomic coherence for the first atomic ensemble.

\section{Supplemental Results \label{sec:supp}}

In this section, we provide extra results to complement those in the main text.

\begin{figure}[!htp]
\begin{centering}
\includegraphics[scale=0.27]{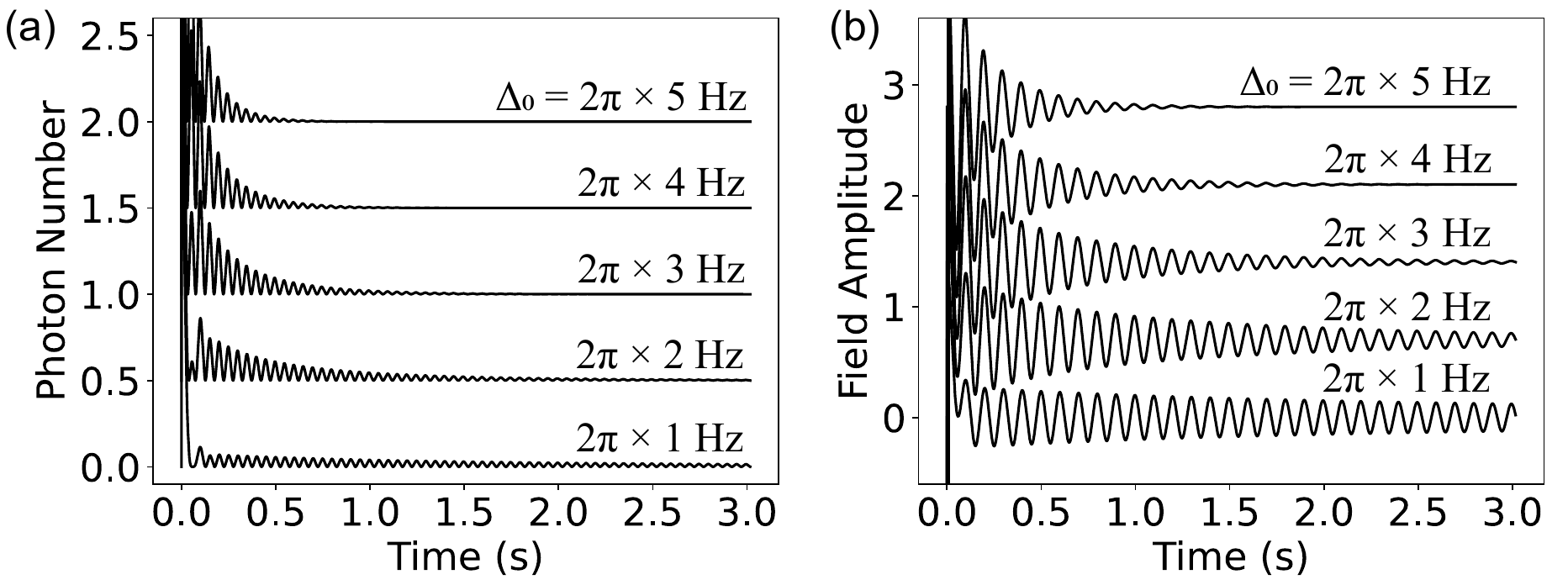}
\par\end{centering}
\caption{\label{fig:S2} Intra-cavity response in the presence of the sinusoidal pulses of the magnetic field $B(t)$, which leads to sinusoidal variation of the frequency detuning $\Delta(t)=\Delta_0 {\rm sin} (\omega_m t)$ of the atomic sub-ensembles to the cavity mode (with $\omega_m=2\pi\times 10$ Hz). Panel (a) and (b) show the intra-cavity photon number (a) and field amplitude (b) as in Fig. 3(d) for $\Delta_0/2\pi=5,4,3,2,1$ Hz (upper to lower curves), where the curves are shifted vertically for the sake of clarity.  }
\end{figure}

In Fig. \ref{fig:S2}, we complement Fig.3(d) in the main text to show how the pulses of magnetic field can be used to control the superradiant pulses. Here, we vary the amplitude of the frequency detuning of the atomic sub-ensembles to the cavity mode from $\Delta_0=2\pi\times 5$ Hz to $2\pi \times 1$ Hz via changing the amplitude of applied magnetic field pulses. We see that as the amplitude $\Delta_0$ decreases from $2\pi \times 5$ Hz to $2\pi \times 1$ Hz, the superradiant signals become weaker at earlier times but persist for a longer duration.

\begin{figure}[!htp]
\begin{centering}
\includegraphics[scale=0.27]{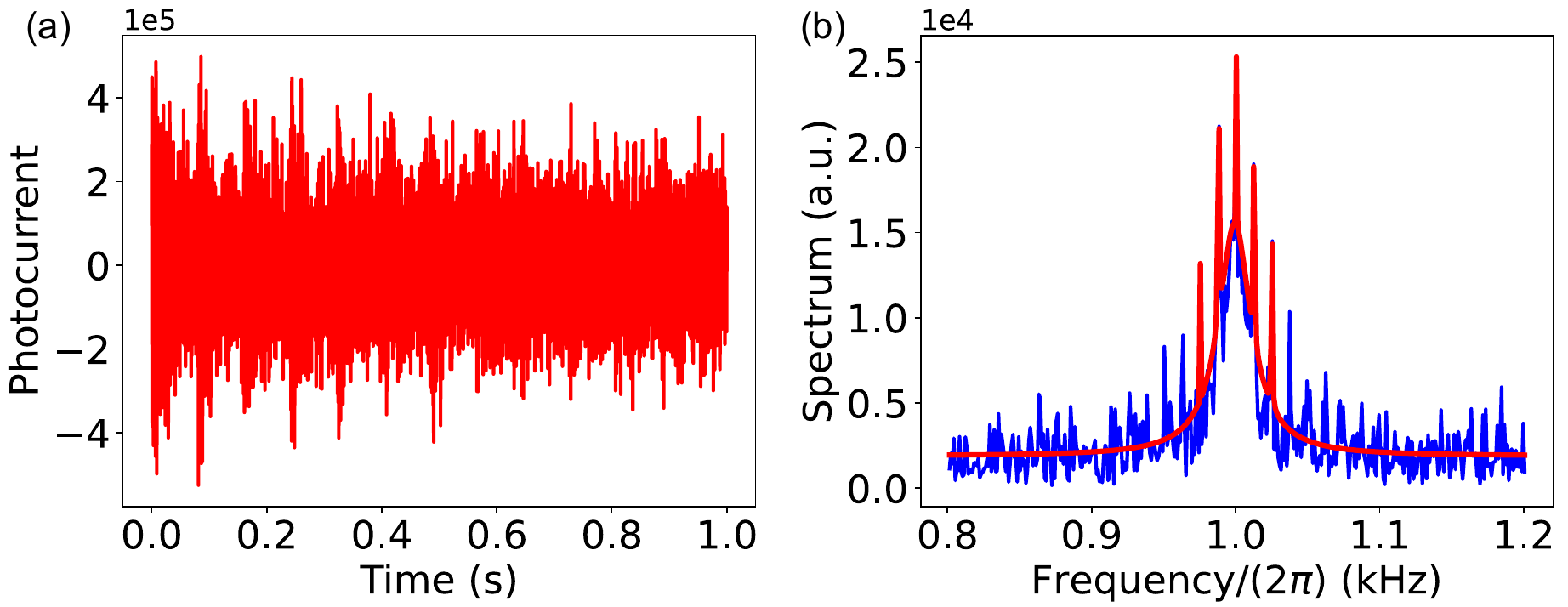}
\par\end{centering}
\caption{\label{fig:S3} Photocurrent (a) and the power spectrum (b) of the prolonged superradiant signal as presented in Fig. 3(b) in the main text. The red curve is the fitting with Lorentzian function (red solid line). }
\end{figure}

Figure \ref{fig:S3} shows the photocurrent (a) and the power spectrum (b) for the superradiant pulses given in Fig. 3(b) of the main text, engineered through rectangular magnetic field pulses. The photocurrent shows spikes with reduced amplitude over the noisy background, which coincides with the change of the cavity field amplitude dips as shown in Fig. 3(b). The spectrum shows also spikes over the noise background in the frequency domain. Although these spikes can be fitted with several Lorentzian functions to determine the frequencies, it is quite difficult to fit them all in a proper way. Thus, in the main text, we have also studied the system response to a smooth sinusoidal magnetic field, and demonstrate that it leads to a power spectrum with two well isolated peaks [see Fig. 4(d) of the main text].

\begin{figure}[!htp]
\begin{centering}
\includegraphics[scale=0.26]{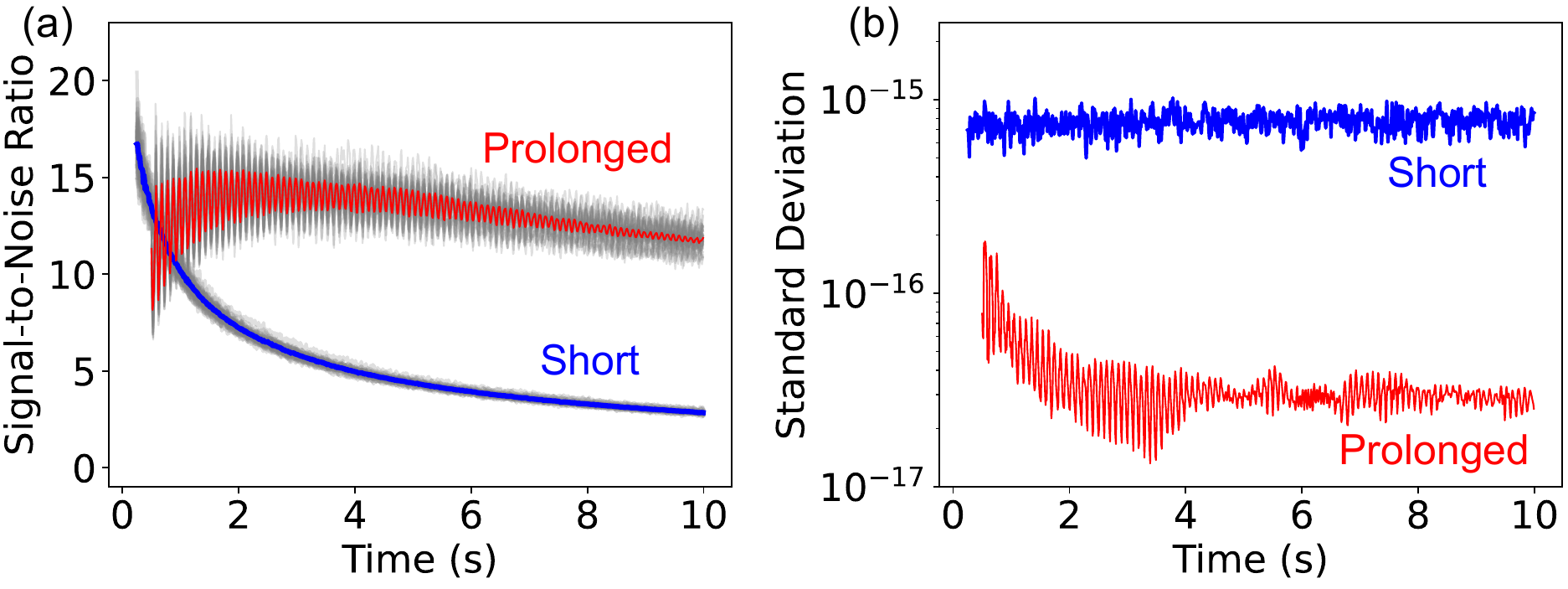}
\par\end{centering}
\caption{\label{fig:S4} Signal-to-noise ratio (a) and  standard deviation of FFD (b) as a function of measurement time $\tau$, obtained using short superradiant pulses (blue curve) and extended pulses (red curve).}
\end{figure}

\begin{figure}[!htp]
\begin{centering}
\includegraphics[scale=0.25]{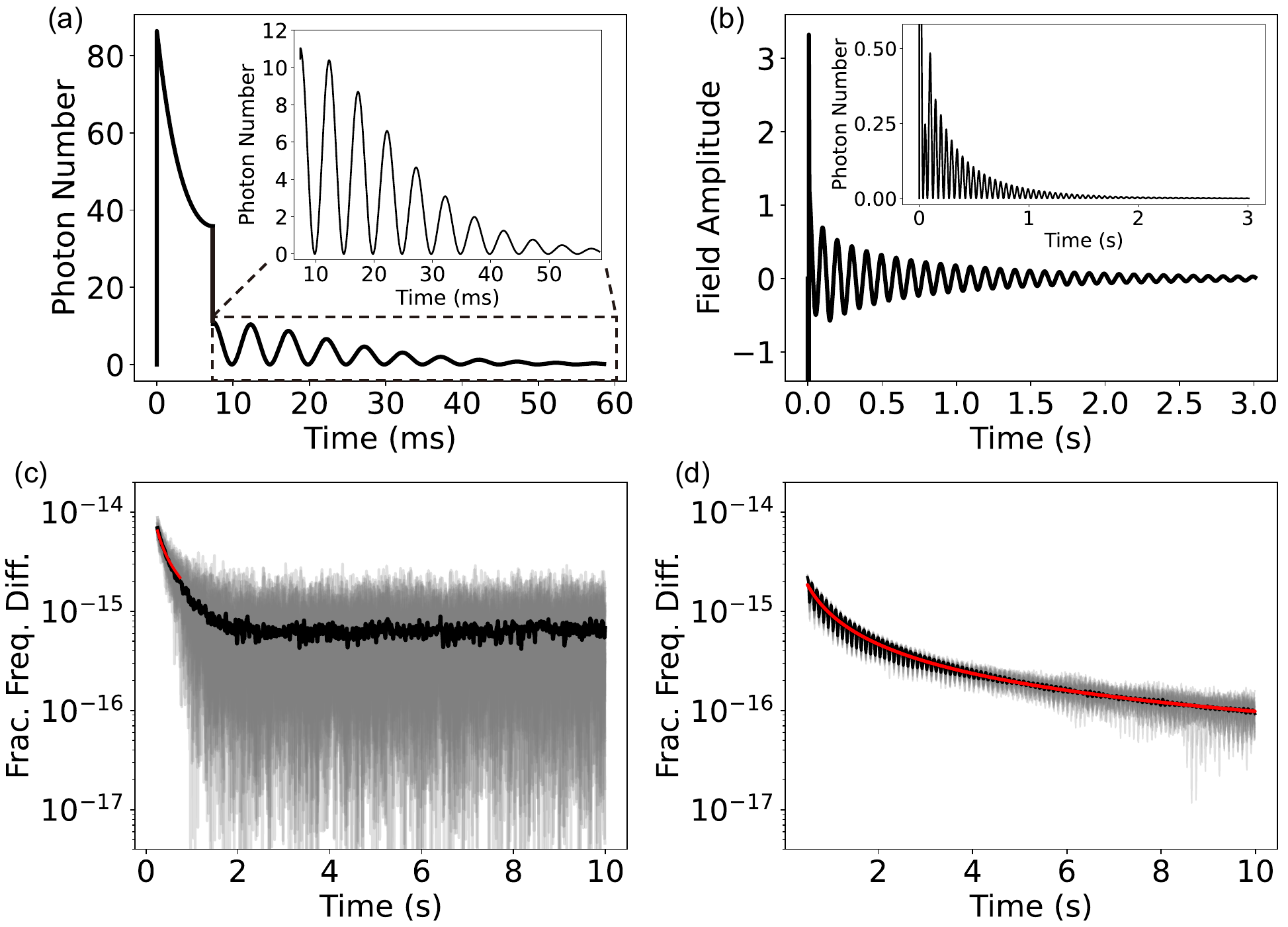}
\par\end{centering}
\caption{\label{fig:S5} Simulations for the system in the presence of the atomic decay and decoherence with the rates $\gamma = 2\pi \times 1$ mHz~\citep{JDolde} and $\chi=2\pi\times 5$ mHz~\citep{SMa2025}.  Panels (a-d) show the same results  as Fig. 2(b), Fig. 3(d), Fig. 4(c) and Fig. 4(f) in the main text, respectively. In the panel (c) and (d), the fits are $1.62\times10^{-15}/(\tau/s)$ and $9.32\times10^{-16}/(\tau/s)$, respectively. }
\end{figure}

Figure \ref{fig:S4} compares further the power spectrum of the short and elongated superradiance through the aspects of the signal-to-noise ratio (SNR) and the standard deviation of the FFD. Here, SNR is defined with the intensity ratio of the peak and the background in the power spectrum. For short pulses, the superradiant signal lasts only approximately 50 ms, leading to a monotonic decrease in SNR. In contrast, prolonged pulses have relatively high intensity during the first 2 s, resulting in a moderate enhancement of SNR within this interval. Beyond 2 s, noise becomes dominant and drives the SNR down, yet the overall SNR remains higher than that of short pulses. The standard deviation of the FFD remains constant at $7.8\times 10^{-16}$ for short pulses. For the prolonged pulses, it decreases slightly within the first 4 s and stabilizes at $3.0\times10^{-17}$ after 4 s, demonstrating a 26 times improvement in frequency measurement stability.

Since the superradiant pulse about $10$ s in our simulations is about one order of magnitude smaller than the lifetime about $150$ s of the upper state of the strontium-87 optical clock transition, we have ignored the possible spontaneous emission and dephasing of such a transition with the rates $\gamma=2\pi \times 1$ mHz~\citep{JDolde} and $\chi=2\pi \times 5$ mHz~\citep{SMa2025} in our studies. To evaluate the quantitative influence of these processes on our proposal, we have extended the conditional master equation~\eqref{eq:cme} by adding the extra Lindblad terms:
\begin{align}
& \partial_t\hat{\rho}_{J}\sim  -\frac{\gamma}{2}\left(\hat{\sigma}^{21}\hat{\sigma}^{12}\hat{\rho}_{J}+\hat{\rho}_{J}\hat{\sigma}^{21}\hat{\sigma}^{12}-2\hat{\sigma}^{12}\hat{\rho}_{J}\hat{\sigma}^{21}\right)\nonumber \\
 & -\chi\left(\hat{\sigma}^{22}\hat{\rho}_{J}+\hat{\rho}_{J}\hat{\sigma}^{22}-2\hat{\sigma}^{22}\hat{\rho}_{J}\hat{\sigma}^{22}\right),\label{eq:cme}
\end{align}
and then modified the programming codes. Based on the updated program, we have recalculated Fig. 2(b), Fig. 3(d), Fig. 4(c), and Fig. 4(f) of the main text, see  Fig.~\ref{fig:S5}, and have not observed significant differences. To analyze further the data, we fit the averaged FFD as a function of the measurement time shown in Fig.~\ref{fig:S5}(c) and (d) with the same expressions as used in the main text. For the frequency measurement with superradiant pulse and prolonged superradiance, we obtain a fitting expression $1.62\times10^{-15}/(\tau/s)$ and $9.32\times10^{-16}/(\tau/s)$, which is comparable to the result $1.69\times10^{-15}/(\tau/s)$ and $9.27\times 10^{-16}/(\tau/s)$ reported in the main text.These results suggest that the decay and dephasing of the optical clock transition have negligible influence on the frequency measurement based on the superradiance.

\end{document}